\documentclass[aps,twocolumn,superscriptaddress,showpacs]{revtex4-1} 
\usepackage{graphicx}
\usepackage{graphicx,epstopdf,color}
\usepackage{amsfonts}
\usepackage{amsmath,amssymb,mathrsfs}
\usepackage{bm}
\usepackage{pst-grad}
\usepackage{nicefrac}

\usepackage{dcolumn}

\def \basp{BaFe$_2$(As$_{1-x}$P$_{x}$)$_2$}

\def \rc{$\rho_c$}
\def \ra{$\rho_{ab}$}

\begin{document}

\title{Anisotropy in the Magnetoresistance Scaling of \basp}

\author{Ian M. Hayes}
\altaffiliation{Contact for correspondence, imhayes@berkeley.edu or analytis@berkeley.edu}
\affiliation{Department of Physics, University of California, Berkeley, California 94720, USA}
\affiliation{Materials Science Division, Lawrence Berkeley National Laboratory, Berkeley, California 94720, USA}

\author{Zeyu Hao}
\affiliation{Department of Physics, University of California, Berkeley, California 94720, USA}

\author{Nikola Maksimovic}
\affiliation{Department of Physics, University of California, Berkeley, California 94720, USA}
\affiliation{Materials Science Division, Lawrence Berkeley National Laboratory, Berkeley, California 94720, USA}

\author{Sylvia K. Lewin}
\affiliation{Department of Physics, University of California, Berkeley, California 94720, USA}
\affiliation{Materials Science Division, Lawrence Berkeley National Laboratory, Berkeley, California 94720, USA}

\author{Mun K. Chan}
\affiliation{Los Alamos National Laboratory, Los Alamos, NM 87545, USA}

\author{Ross D. McDonald}
\affiliation{Los Alamos National Laboratory, Los Alamos, NM 87545, USA}

\author{B. J. Ramshaw}
\affiliation{Los Alamos National Laboratory, Los Alamos, NM 87545, USA}
\affiliation{Laboratory of Atomic and Solid State Physics, Cornell University, Ithaca, NY, 14853.}

\author{Joel E. Moore}
\affiliation{Department of Physics, University of California, Berkeley, California 94720, USA}
\affiliation{Materials Science Division, Lawrence Berkeley National Laboratory, Berkeley, California 94720, USA}

\author{James G. Analytis}
\altaffiliation{Contact for correspondence, imhayes@berkeley.edu or analytis@berkeley.edu}
\affiliation{Department of Physics, University of California, Berkeley, California 94720, USA}
\affiliation{Materials Science Division, Lawrence Berkeley National Laboratory, Berkeley, California 94720, USA}

\begin{abstract}
Theories of the strange metal, the parent state of many high temperature superconductors, invariably involve an important role for correlations in the spin and charge degrees of freedom. The most distinctive signature of this state in the charge transport sector is a resistance that varies linearly in temperature, but this phenomenon does not clearly point to one mechanism as temperature is a scalar quantity that influences every possible mechanism for momentum relaxation. In a previous work we identified an unusual scaling relationship between magnetic field and temperature in the in-plane resistivity of the unconventional superconductor \basp, providing an opportunity to use the vector nature of the magnetic field to acquire additional clues about the mechanisms responsible for scattering in the strange metal state. Here we extend this work by investigating other components of the conductivity tensor under different orientations of the magnetic field. We find that the scaling phenomenon involves only the out-of-plane component of the magnetic field and is, strikingly, independent of the direction of the applied current. This suggests that the origin of the strange magnetotransport is in the action of the magnetic field on the correlated behavior of spin and charge degrees of freedom, rather than on the simple cyclotron motion of individual quasiparticles.
\end{abstract}
\pacs{74.25.Dw, 74.25.fc, 74.40.Kb}

\maketitle

The formidable task of understanding high temperature superconductivity is tied to understanding the so-called strange metal state which exists near optimal doping. This state is common to both the copper and iron based superconductors, and it has several accepted experimental signatures~\cite{stewart_non-fermi-liquid_2001}; in the charge transport sector these include a $T$-linear resistivity down to low temperatures and a temperature dependent Hall effect. There is not, however, any widely accepted theory for the strange metal state. Although it is generally agreed that a successful theory will involve some physics beyond the standard theory of metals, Landau's theory of the Fermi liquid \cite{landau_FL_1957}, ideas have varied widely about what that change ought to be.~\cite{anderson_hidden_2011, varma_fluctuations_2016, sachdev_fluctuating_2009, hartnoll_universal_2014, dalidovich_nonlinear_2004, green_nonlinear_2005, davison_holographic_2014} In this environment, experiments that can identify general features of the new physics, and especially broad features of the dominant scattering mechanisms, are invaluable.

The magnetic field dependence of the resistivity (MR) has occasionally been included as part of the strange metal phenomenology because the MR in these compounds is known to violate the scaling relationship between $H$ and $\rho(H = 0)$, known as Kohler's rule, that is expected in a simple metal~\cite{harris_violation_1995}~\cite{kasahara_evolution_2010}. However, even within a quasiparticle picture there are many reasons why Kohler's rule could be violated, including the existence of multiple scattering times on different Fermi surface sheets, or of a temperature or wavelength dependence of the scattering objects, including phonons. In any case, no single, simple pattern has been observed in the MR that could guide theoretical efforts. 
Recently, we reported an anomalous scaling relationship between magnetic field and temperature in the MR of the unconventional, high-$T_c$ iron-pnictide superconductor \basp\ near optimal doping.\cite{hayes_scaling_2016} The existence of scaling between $H$ and $T$ in the MR is intriguing both because it is difficult to obtain in a Fermi liquid picture and because it connects the physics of $T$-linear resistivity to the effects of a magnetic field, providing a new window into the strange metal state. In particular, it was found that field has the same scaling dimension as temperature, leading to an $H-$linear MR at low temperatures with a gradient that is related to the gradient in temperature by the simple ratio of fundamental constants $\mu_B/k_B$. Since this is exactly the energy scale associated with a single electron spin in a magnetic field, it is a natural hypothesis that this arises from a single spin effect. 

\begin{figure*}[ht]
\includegraphics[width=18cm]{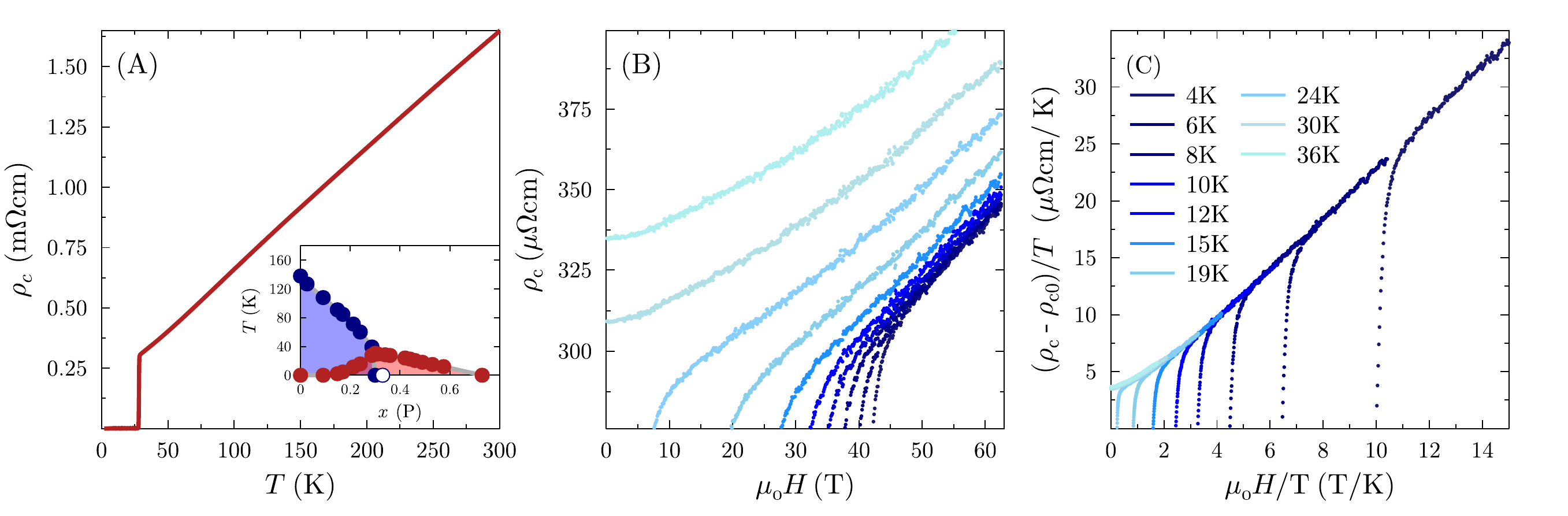} 
\caption{{\bf Scaling in $\rho_c$ of \basp.} {\bf A.} The interlayer resistivity, \rc, as a function of temperature. The inset shows a schematic phase diagram of \basp, showing the AFM and superconducting transitions, and the location of optimal doping (white circle). {\bf B.} Interlayer resistivity as a function of magnetic field up to 63T at temperatures ranging from 4K to 36K. The $H-$linear MR is apparent at low Temperatures where $\rho$ is roughly independent of $T$. {\bf C.} A scaling plot of the MR curves shown in {\bf B}. The residual resistivity (found by fitting the 4K curve to a line and taking the $H=0$ intercept) is subtracted off and the remainder is normalized to the temperature and plotted versus $H/T$.}
\label{fig:Scaling} 
\end{figure*}

In this Letter, we extend our analysis of the high-field MR of \basp\ to other orientations of the field and current. We find that the interlayer resistivity, \rc , shows the same scaling behavior as \ra, but only when the field is oriented along the same crystallographic axis. Furthermore, despite a factor of five difference in the magnitude of \ra\ and \rc, the fractional change in the two resistivities is very nearly the same as a function of field. In contrast, neither \ra\ nor \rc\ shows scaling in $H/T$ when the field lies in the plane. Measurements of the angle dependence of the MR show that it is specifically the component of the field along the $c-$axis that enters the scaling expression. 
These observations point to a scenario in which the MR arises not from the action of the magnetic field on the orbital or spin degrees of freedom of the individual quasiparticles, but from the coupling of the field to some collective dynamics in the material. This behavior provides an important window into the current-relaxation mechanisms that are responsible for the anomalous transport properties mentioned above. It potentially also has relevance for several noteworthy ideas about unconventional quantum dynamics that have emerged in recent years, including an improved understanding of  $\hbar / (k_B T)$ as a fundamental scale,~\cite{maldacena_bound_2016}  and studies~\cite{banerjee_scrambling_2017, balents_StronglyCorrelated_2017} that have based theories of unconventional transport on the crossover from the non-Fermi-liquid behavior in the Sachdev-Ye-Kitaev model~\cite{sachdev_SYK_1993} to a Fermi liquid when more conventional hopping terms are added. 

Single crystals of \basp\ were grown by a self-flux method described elsewhere \cite{analytis_enhanced_2010}. The phosphorous content of these materials was previously determined using X-ray photoelectron spectroscopy. Samples for this study were taken from the same or similar batches and found to have the anticipated $T_c$, which correlates well with the phosphorous fraction, $x$. For \rc\ measurements, small, plate-like like crystals with typical dimensions of $150 \mu m \times 150 \mu m \times (50 - 90 )\mu m$ ($a \times b \times c$) were contacted by covering each $ab$-face with tin-lead solder, as was done in previous studies of the interlayer resistivity in these materials~\cite{tanatar_interlayer_2013}. These contacts had very small (10-50 $\mu \Omega$) resistances, which represent a small fraction ($~1\%$) of the sample signal. This allowed us to measure \rc\ by performing a four-point, AC lock-in resistance measurement of the contact-sample-contact system and neglecting the contribution from the contacts. For \ra\ measurements, a standard four-point measurement was used, with contacts made by first sputtering gold onto the contact areas and then attaching 25 $\mu$m gold wires with Epotek H20e silver epoxy. 
Measurements were performed in pulsed magnetic fields of up to 65 Tesla at the NHMFL Pulsed Field Facility, Los Alamos National Laboratory. 

For the interlayer resistivity, \rc, we report measurements of the MR near optimal doping ($x\sim0.3$) for both the longitudinal ($H$ parallel to $c$) and transverse ($H$ parallel to $a$) configurations. Figure \ref{fig:Scaling} shows the data for the first of these configurations, from which two facts are immediately apparent. First, the MR is linear in field at the lowest temperatures and gradually develops curvature at higher temperatures and especially at lower fields. Second, it is almost temperature independent for low temperatures and high fields, as if the value of the magnetic field were the only important variable for determining the resistivity. These are the two most visible consequences of the scaling form found for the resistivity of \ra \cite{hayes_scaling_2016}. Since the resistivity is linear in field at low temperatures, we can extrapolate that curve to zero field to obtain the residual resistivity and perform a scaling analysis on the temperature/field dependent part of the resistivity. Panel C of Figure \ref{fig:Scaling} shows the results of this analysis, where the data collapse onto a single, hyperbola-like curve. 

\begin{figure}[ht]
\includegraphics[width=8.5cm]{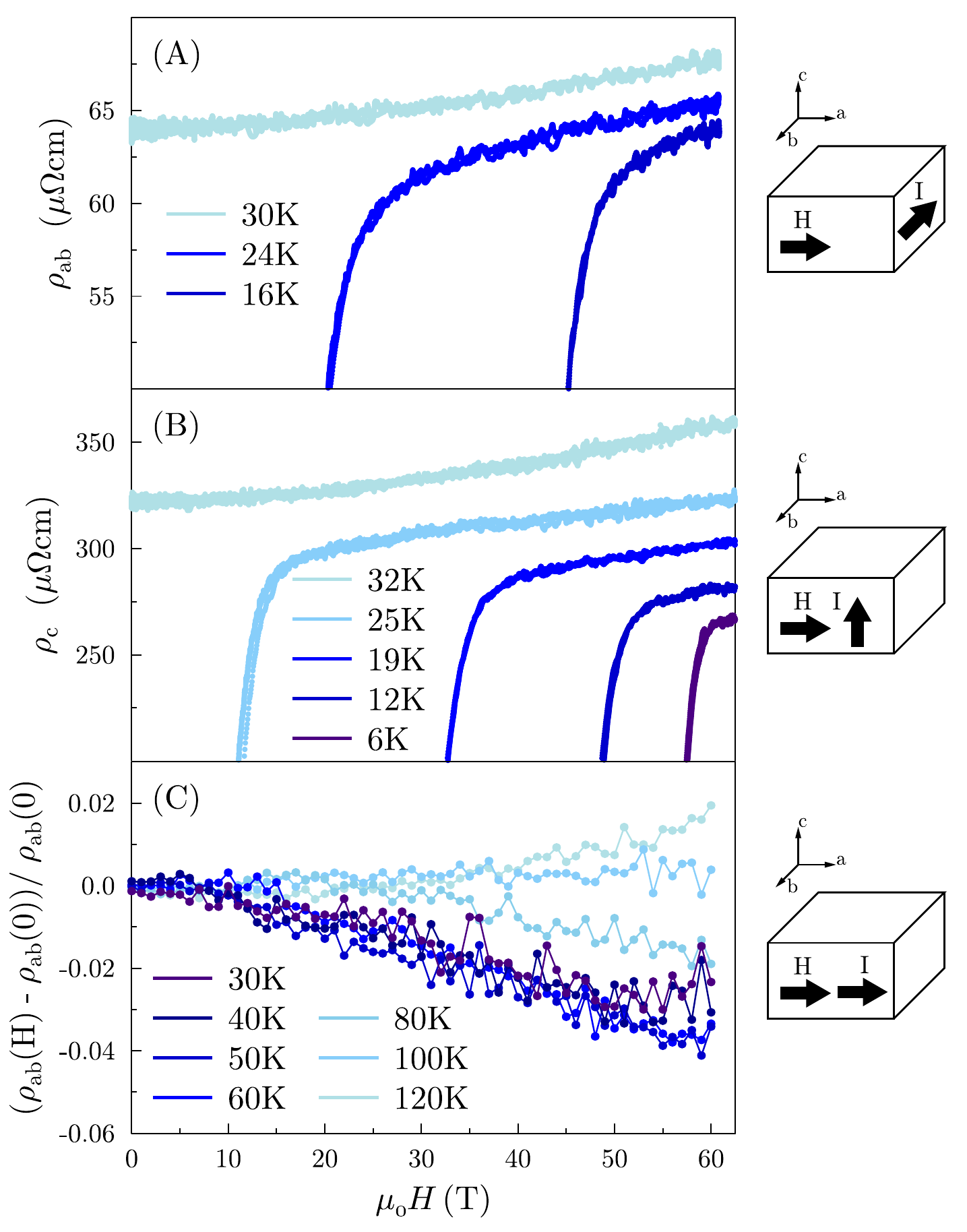} 
\caption{{\bf MR of \basp\ in \ra\ and \rc\ with $H$ in the plane.} {\bf A.} MR of \ra\ as a function of transverse (in-plane) magnetic field up to 63T at temperatures ranging from 4K to 36K. Although $H_{C2}$ is very close to 60 Tesla, we can see that there is still strong temperature dependence at high fields, and thus $H-T$ scaling of the form found with field along the c-axis is not present. {\bf B.} The same plot but for \rc. {\bf C.} Fractional MR of \ra\ in the longitudinal configuration. Below about 100K the MR becomes negative and has a similar magnitude to the transverse, in-plane configuration. Diagrams on the right are schematic representations of each experimental configuration.}
\label{fig:H_in_Plane_MR} 
\end{figure}

The collapse of the data observed in Figure \ref{fig:Scaling} shows that \rc\ has the same qualitative behavior in $H$ and $T$ as the in-plane resistivity, \ra~\cite{hayes_scaling_2016}. The quantitative details of the data are similar as well. First, considered as a fraction of the zero field resistivity, the MR is of the same scale in \rc\ and \ra\ at low temperatures.  Below 30K there is no normal state resistivity in zero field to compare to, but at thirty Kelvin the MR of \rc\ at sixty Tesla is $27\%$, while for the same values of $H$ and $T$ it is $36\%$ in \ra~\cite{hayes_scaling_2016}. This similarity is especially significant given that \rc\ is about five times \ra.
The other important quantity that goes into the  MR scaling is the scale factor connecting $H$ and $T$. This can be found by comparing the $H/T = 0$ intercept of the hyperbola in Figure \ref{fig:Scaling} (C) to the slope of the hyperbola at high values of $H/T$.  In \ra\ this quantity is found to be equal to $\mu_B/k_B$ to within a few percent~\cite{hayes_scaling_2016}. In \rc\ this quantity is found to be about 5$\%$ smaller, but this difference is within the error of the fit. Taken together, these data show an interlayer magnetoresistance nearly identical to that seen in \ra.
 
It is highly unusual for the MR to be independent of current direction~\cite{pippard_magnetoresistance_2009}. In an ordinary Fermi Liquid, the effect of a magnetic field on the resistivity comes principally from the action of the Lorentz force on the current-carrying quasiparticles. Under the influence of an electric field, the Fermi sea shifts and a net current comes from the excess of quasiparticles with momentum in one direction. If a magnetic field is also applied, the excited quasiparticles are deflected around the Fermi surface, reducing their average velocity in the direction of current flow, and therefore reducing the corresponding conductivity. Because this reduction of the quasiparticle velocity depends on the sign and magnitude of the curvature of the Fermi surface, and because there is a different component of the Fermi velocity that is relevant for \ra\ and \rc, this process is usually exquisitely sensitive to the relative angles of the field, current and the crystallographic axes. Indeed, the angular dependence of the MR has historically been a powerful way to map the shape of the Fermi surface in ordinary metals, even without the observation of Subinikov-de Haas oscillations~\cite{pippard_magnetoresistance_2009}. The indifference of the MR scaling to current direction suggests an origin which is more complex than the cyclotron motion of quasiparticles under the action of the Lorentz force.

This is the first major finding of this study, and it is intriguing because it supports the idea that the dynamics of quasiparticles do not dominate the transport properties of these materials, especially in their transport properties. It also naturally raises the question of whether the scaling is independent of field orientation as well. There are three further field-current-crystal orientations to consider: the longitudinal MR of \ra\ (Figure 2 C), the transverse MR of \ra\ with field in the plane (Figure 2 A), and the transverse MR of \rc\ (Figure 2 B). There is only one transverse MR for \rc\ because \basp\ is tetragonal at optimal doping. With field along the $a-$axis $H_{C2}$ is around sixty Tesla, leaving almost no normal state accessible at the lowest temperatures. However, since the object of interest in these measurements is a scaling relation in $H$ and $T$, it should still be observable in a sixty-five Tesla magnet.
 
The two configurations of transverse MR with field in the plane show broadly similar behavior, and show clear differences with the MR for $H$ aligned along the $c-$axis (Figure 2 A, B). First, the overall scale of the MR is much smaller: ~4$\%$ at 30 Kelvin and 60 Tesla. Second, although there is not enough normal state at low temperatures to make a definitive statement about linearity, it is clear that the resistivity is not independent of temperature at high fields, which immediately tells us that the data do not scale in the same manner as those taken with the field along the $c-$axis. Scaling analyses confirm this intuition. Therefore $H/T$ scaling seems to be a special property of the charge transport when the magnetic field is normal to the iron-arsenide plane.

\begin{figure}[ht]
\includegraphics[width=8.5cm]{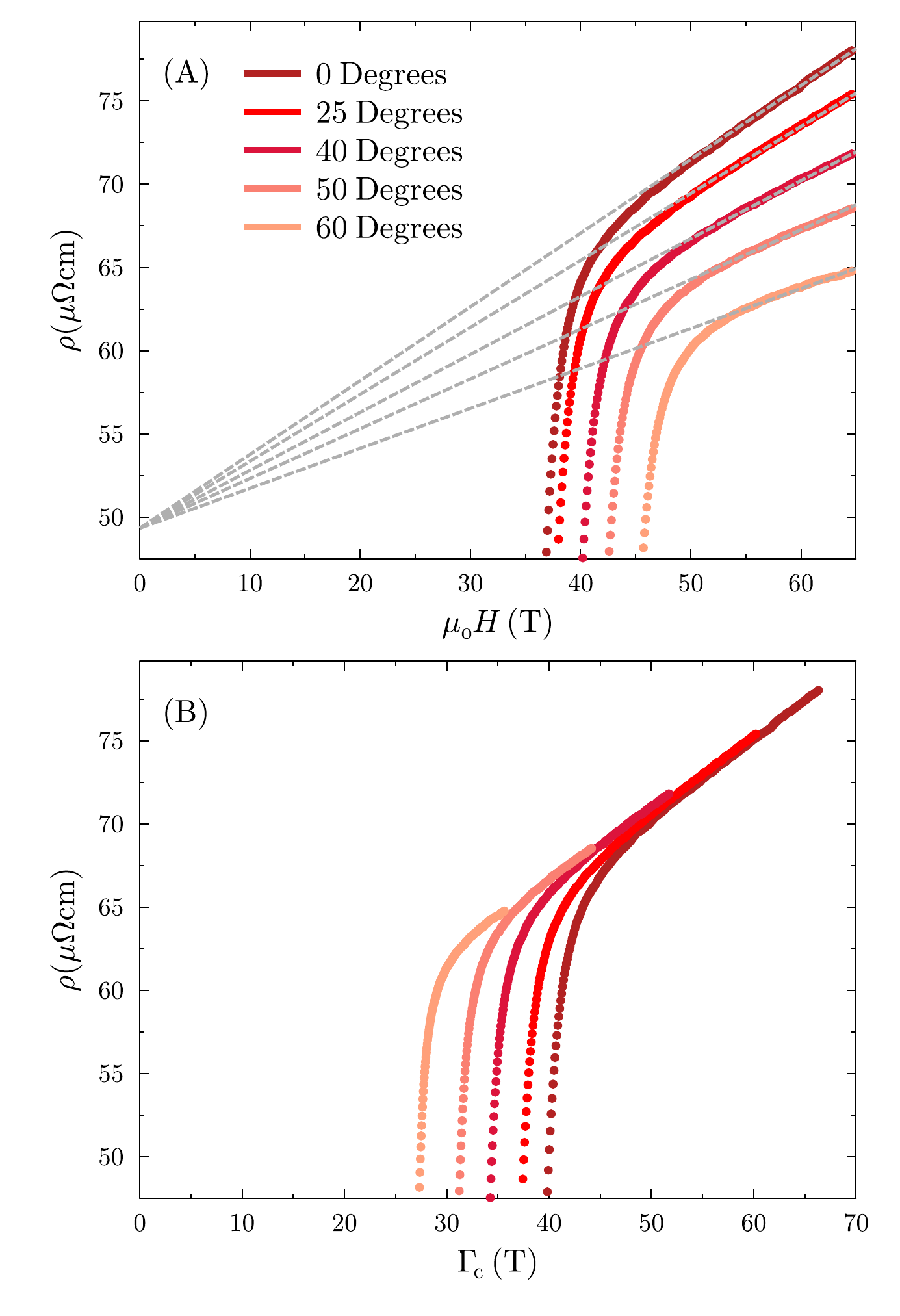} 
\caption{{\bf Transverse MR of \ra\ and with $H$ rotated away from the $c-$axis.} {\bf A.} Raw MR as a function of $H$ for several orientations of the magnetic field. The field was always maintained perpendicular to the current but rotated away from the $c-$axis. The temperature is 10K. {\bf B.} The same curves plotted in panel (A), but this time as a function of the combined field-temperature energy scale, $\Gamma_c$, using only the component of the field along the $c-$axis.}
\label{fig:Rotation_Scaling} 
\end{figure}

To test whether the $c-$axis component of the magnetic field really is the only relevant part for the scaling, we measured the angle dependence of \ra. Figure \ref{fig:Rotation_Scaling} shows the in plane resistivity as a function of $H$ as the field is rotated from the out-of-plane orientation to the in-plane orientation, all the while remaining orthogonal to the current. 
These data were taken at ten Kelvin; this temperature was chosen to be low enough for the MR to be $H-$linear, but not so low that the upper critical field would prevent us from measuring anything for large angles. The $H-$linear MR is present for small deviations from the scaling configuration, but its slope decreases at higher angles. Significantly, the extrapolated intercepts of these low-angle curves are all the same (Figure \ref{fig:Rotation_Scaling} A), suggesting that we are seeing the same $H-$linear MR, but with a smaller effective field. To see this, we can plot these curves against the quadrature sum of $H_c = H cos(\theta)$ and $T$, $\Gamma_c = \sqrt{(k_BT)^2 + \mu_B\mu_0H_c^2}$. As shown in Figure \ref{fig:Rotation_Scaling} (B) this causes a remarkable collapse of the data at low angles. The scaling in the MR therefore follows the $c$-axis component of magnetic field and only this component.

There is one final configuration of field and current: longitudinal MR in the plane. Interestingly, the MR in this configuration is negative with a non-negligible magnitude of about 4$\%$ (comparable in magnitude to the positive MR in the transverse, in-plane configuration). The negative longitudinal magnetoresistance (LMR) is strongly temperature dependent, turning on around 100K, approximately where the resistivity starts to deviate from strict $T-$linear behavior (Figure \ref{fig:Scaling} A). Negative MR in a metal with a large Fermi surface is highly unusual, even in the longitudinal configuration. General considerations lead us to expect that an applied magnetic field will increase the resistivity since it will deflect quasiparticles away from the direction of the current~\cite{pippard_longitudinal_1964}. A negative MR can occur naturally, however, if the dominant scattering objects are magnetic, since one effect of an applied magnetic field could easily be to reduce the scattering from those objects. 

The facts reported in this letter show that the magnetoresistance near optimal doping in \basp\ is not driven by single particle dynamics. The observation that the scaling only appears for magnetic field along the $c$-axis suggests that the effect is not from coupling to individual spins, as such a strong anisotropy in Zeeman coupling is unlikely. Additionally, the observation of the scaling for both in-plane and $c$-axis currents rules out most straightforward cyclotron effects, as those are generally sensitive to the current direction. As the scaling is observed on the disordered side of a phase transition to in-plane AFM order (inset of Figure \ref{fig:Scaling} A), it seems probable that the it is related to magnetic fluctuations of this in-plane order, which are naturally more sensitive to transverse (i.e., $c$-axis) fields. In the ordered state, it is known that the magnetic response is highly anisotropic~\cite{chu_inplane_2010}. Hence the coupling of conduction electrons to collective magnetic fluctuations on a variety of length scales should be included in models of the observed scaling behavior. This emerges naturally in approaches to the strange metal state that are based on critical fluctuations of an order parameter, including the marginal Fermi liquid theory of Varma and collaborators~\cite{varma_phenomenology_1989, varma_fluctuations_2016}. Whether or not the naive picture of proximate AFM fluctuations is relevant, the observation about the quasi-2D nature of the field-temperature scaling in the transport should serve as a valuable guide for model building, and should help distinguish between among various theoretical scenarios that attempt to explain the $T-$linear resistivity of the strange metal.

We thank Chandra Varma, Philip Phillips, Aavishkar Patel and Subir Sachdev for fruitful discussions. This work is supported by the Gordon and Betty Moore Foundation's EPiQS Initiative through Grant GBMF4374. A portion work was performed at the National High Magnetic Field Laboratory, which is supported by National Science Foundation Cooperative Agreement No. DMR-1157490 and DMR-1644779 and the State of Florida. B.R., M-K.C. and R.D.M. acknowledge funding from the U.S. Department of Energy Office of Basic Energy Sciences Science at 100 T program. S.K.L. acknowledges support from the National Science Foundation Graduate Research Fellowship under Grant No. DGE 1106400. J. E. M. acknowledges the support of the Quantum Materials program at Lawrence Berkeley National Laboratory, which is supported by the Director, Office of Science, Office of Basic Energy Sciences, Materials Sciences and Engineering Division, of the U.S. Department of Energy under Contract No. DE-AC02-05CH11231. 



\end{document}